\documentclass{aipproc}

\newcommand{\be}{\begin{equation}}
\newcommand{\ee}{\end{equation}}
\newcommand{\bea}{\begin{eqnarray}}
\newcommand{\eea}{\end{eqnarray}}

\layoutstyle{6x9}

\begin{document}

\def\slash#1{#1\!\!\!/}

\title
      [Crystalline Color Superconductivity]
      {Crystalline Color Superconductivity}


\author{Krishna Rajagopal}{
  address={Center for Theoretical Physics, Massachusetts Institute of
Technology, Cambridge,~MA,~USA~02139. E-mail: krishna@ctp.mit.edu},
  email={krishna@ctp.mit.edu},
  thanks={Research supported in 
part  by the DOE under cooperative research agreement \#DF-FC02-94ER40818
and through an OJI Award, and by the
A. P. Sloan Foundation.}
}

\begin{abstract}
We give an introduction crystalline color superconductivity,
arguing that it is likely to occur wherever quark matter 
in which color-flavor locking does not occur is found.
We survey the properties
of this form of quark matter, and argue that its presence
in a compact star may result in pulsar glitches, and thus
in observable consequences.  However, elucidation of this proposal
requires an understanding of the crystal structure, 
which is not yet in hand.
\end{abstract}

\date{\today}

\maketitle

\section{Introduction}

At asymptotic densities, the ground state of QCD with three
quarks with equal masses is expected to be the color-flavor locked (CFL) 
phase \cite{CFL,OtherCFL,reviews}.
This phase features a condensate of Cooper pairs of
quarks which includes $ud$, $us$, and $ds$ pairs. Quarks
of all colors and all flavors participate in the
pairing, and all excitations with quark quantum numbers are
gapped. 

The CFL phase persists for unequal quark masses, so long as the 
differences are not too large \cite{ABR2+1,SW2}.
It is very likely the ground
state for real QCD, assumed to be in equilibrium
with respect to the weak interactions, over a substantial
range of densities.  In this phase, chiral symmetry is
broken via the locking of left-flavor and right-flavor
symmetries to color~\cite{CFL}.  Terms of order $m_s^4$
in the effective Lagrangian for the resulting pseudo-Goldstone
bosons~\cite{Casalbuoni} 
may rotate the CFL condensate in the $K^0$ 
direction~\cite{BedaqueSchaefer}, resulting in
further pseudo-Goldstone bosons~\cite{K0condensateGB}.
Throughout the range of parameters
over which the CFL phase exists as a bulk (and therefore
electrically neutral) phase, it consists of equal
numbers of $u$, $d$ and $s$ quarks and is therefore
electrically neutral in the absence of any 
electrons \cite{neutrality}.  
The equality of the three quark number densities is 
enforced in the CFL phase
by the fact that this equality maximizes the pairing energy
associated with the formation of $ud$, $us$, and $ds$ Cooper pairs.
This equality is enforced even though the strange quark, with mass $m_s$,
is heavier than the light quarks.  
If higher order effects do in fact introduce an additional $K^0$ 
condensate, the conclusion that the CFL phase is electrically neutral in the
absence of electrons remains, because $K^0$ mesons are neutral.

If one imagines increasing $m_s$ (or, more physically, decreasing $\mu$)
color-flavor locking is maintained until a transition
to a state of quark matter in which 
some quarks remain ungapped.  This ``unlocking transition'', which  
must be first order~\cite{ABR2+1,SW2}, occurs when
$m_s^2\approx 4\mu\Delta_0$~\cite{neutrality,ARRW}.
In this expression, $\Delta_0$ is the BCS pairing gap,
estimated in both models and asymptotic analyses to be of order 
tens to 100 MeV~\cite{reviews}.  The strange quark mass $m_s$ 
is a density-dependent effective mass~\cite{BuballaOertel}.  
For $\mu\sim 400-500$ MeV,
corresponding to quark matter at densities which may arise
at the center of compact stars, $m_s$ is certainly significantly
larger than the current quark mass, and its value is not known.
In drawing the QCD phase diagram, therefore, there are two
possibilities.  As a function of decreasing $\mu$, one
possibility is a first order phase transition directly from
color-flavor locked quark matter to hadronic matter,
as explored in Ref. \cite{ARRW}.  The second possibility
is an unlocking transition~\cite{ABR2+1,SW2} to quark matter
in which not all quarks participate in the dominant pairing,
followed only at a lower $\mu$ by a transition to hadronic matter.
We assume the second possibility here, and explore its consequences.

In quark matter in which CFL pairing
involving all quarks does not occur, it is likely 
that up and down quarks continue to pair.  In this 2SC
phase, which was the earliest color superconducting 
phase to be studied~\cite{2SC},
the attractive
channel involves the formation of Cooper pairs which
are antisymmetric in both color and 
flavor, yielding
a condensate with color (greek indices)
and flavor (latin indices) structure 
$\langle q^\alpha_a  q^\beta_b \rangle\sim \epsilon_{ab}
\epsilon^{\alpha\beta 3}$.  
This condensate leaves five quarks unpaired: up and down quarks
of the third color, and strange quarks of all three colors.
Because the BCS pairing scheme leaves ungapped quarks with
differing Fermi momenta,
it is natural to ask whether there is some generalization
of the pairing ansatz, beyond BCS, in which pairing 
between two species of quarks persists even once their
Fermi momenta differ.   Crystalline color superconductivity
is the answer to this question.

\section{The Crystalline Color Superconducting State}

To date, crystalline color superconductivity has only
been studied in the simplified model context in which
one considers only
pairing between massless up and down quarks whose
Fermi momenta we attempt to push apart
by turning on a chemical
potential difference~\cite{BowersLOFF,ngloff,LOFFphonon,pertloff},
rather than considering CFL pairing in the presence of
quark mass differences.                                            
That is, we introduce 
\begin{eqnarray}\label{mubardmu}
\mu_u&=&\mu-\delta\mu\nonumber\\
\mu_d&=&\mu+\delta\mu\ .
\end{eqnarray}
If $\delta\mu$ is nonzero but less than some $\delta\mu_1$,
the ground state 
is precisely that obtained for 
$\delta\mu =0$ \cite{Clogston,Bedaque,BowersLOFF}.  In this 2SC state,
red and green up and down quarks pair, yielding four quasiparticles
with superconducting gap $\Delta_0$ \cite{2SC}.
Furthermore, the number density of red and green up quarks is the
same as that of red and green down quarks. 
As long as $\delta\mu$ is not too large, this BCS state
remains unchanged (and favored) 
because maintaining equal number densities, and
thus coincident Fermi surfaces, 
maximizes the pairing
and hence the gain in interaction energy.  
As $\delta\mu$ is increased, the BCS state remains the ground state of
the system only as long as its negative interaction
energy offsets the large positive free energy cost 
associated with forcing the Fermi seas to deviate from their
normal state distributions. In the weak coupling limit, in
which $\Delta_0/\mu\ll 1$, the BCS state persists for
$\delta\mu<\delta\mu_1=\Delta_0/\sqrt{2}$ \cite{Clogston,BowersLOFF}.
These conclusions are the same (as long as $\Delta_0/\mu\ll 1$)
whether the interaction between quarks is
modeled as a point-like four-fermion interaction or 
is approximated
by single-gluon 
exchange~\cite{reviews}.
The loss of BCS pairing at $\delta\mu=\delta\mu_1$  is the
analogue in this toy model of the unlocking transition.

If $\delta\mu$ is too large, no pairing between species is
possible. The transition between the BCS and unpaired states as 
$\delta\mu$ increases past $\delta\mu_1$ has been studied in
electron superconductors~\cite{Clogston} 
and QCD
superconductors~\cite{ABR2+1,SW2,Bedaque} assuming 
that no other state intervenes.  However,
there is good reason to think that another state can occur.  This is
the ``LOFF'' state, first explored by Larkin and Ovchinnikov\cite{LO}
and Fulde and Ferrell\cite{FF} in the context of electron
superconductivity in the presence of magnetic impurities.
They found that near the
unpairing transition, in a range $\delta\mu_1<\delta\mu<\delta\mu_2$,
it is favorable to form a state in
which the Cooper pairs have nonzero momentum. 
This generalization of the pairing ansatz (beyond BCS ans\"atze
in which only quarks with momenta which add to zero pair) is favored because
it gives rise to a region of phase space where each of the two quarks
in a pair can be close to its Fermi surface,
and such pairs can be created at low cost in free energy.
Condensates of this sort spontaneously
break translational and rotational invariance, leading to
gaps which vary periodically in a crystalline pattern.
If in some shell within the quark matter core
of a neutron star (or within a strange quark star)  
the quark number densities are
such that crystalline color superconductivity arises,
rotational vortices may be pinned in this shell, making
it a locus for glitch phenomena~\cite{BowersLOFF}.

In Ref.~\cite{BowersLOFF}, 
we have evaluated the width of the window 
$\delta\mu_1<\delta\mu<\delta\mu_2$
for which crystalline color superconductivity occurs, 
upon making the simplifying 
assumption that
quarks interact via a
four-fermion interaction
with the quantum numbers of single gluon exchange.

In the LOFF state, each Cooper pair carries 
momentum $2{\bf q}$ with $|{\bf q}|\approx 1.2 \delta\mu$.
The condensate and gap parameter vary in space with wavelength
$\pi/|{\bf q}|$.  Although the magnitude $|{\bf q}|$ is determined
energetically, as we sketch below, the direction $\hat{\bf q}$
is chosen spontaneously.  
The LOFF state is characterized by a gap parameter $\Delta$ and a 
diquark condensate, but not by an energy gap in the dispersion
relation: we obtain the quasiparticle dispersion 
relations\cite{BowersLOFF} and find that they vary
with the direction of the momentum, yielding gaps that vary from zero
up to a maximum of $\Delta$.  The condensate is dominated by
the regions in momentum space in which a quark pair
with total momentum $2{\bf q}$ has both members of
the pair within $\sim \Delta$ of their respective 
Fermi surfaces. These regions form circular bands
on the two Fermi surfaces.  Choosing a single, fixed, ${\bf q}$
means that only one circular band
on each Fermi surface participates in the pairing.
In all work published to date,
we assume that all Cooper pairs make the same choice of
direction $\hat{\bf q}$.  Making this ansatz corresponds
to choosing a single circular band on each Fermi surface.
In position space, it corresponds to a condensate which
varies in space like 
\begin{equation}\label{simplifiedcondensate}
\langle \psi({\bf x}) \psi({\bf x})\rangle \propto \Delta e^{2i{\bf q}
\cdot {\bf x}}\ .  
\end{equation} 
This ansatz is certainly {\it not} the best choice.
If a single plane wave is favored, why not two? That is,
if one choice of $\hat{\bf q}$ is favored, why not add 
a second ${\bf q}$, with the same $|{\bf q}|$ but
a different $\hat{\bf q}$?  If two are favored, why not three?
This question, namely the determination of the favored crystal
structure of the crystalline color superconductor phase,
remains open.  Note, however, that if we find a region
$\delta\mu_1<\delta\mu<\delta\mu_2$ in which the simple
LOFF ansatz, with a single $\hat{\bf q}$, is favored over
the BCS state and over no pairing, then the LOFF state
with whatever crystal structure turns out to be optimal
must be favored in {\it at least} this region.  Note also
that even the single $\hat{\bf q}$ ansatz, which we use
henceforth, breaks translational and rotational invariance
spontaneously.  The resulting phonon has been analyzed
in considerable detail in Ref. \cite{LOFFphonon}.
It will be very interesting to use these methods to 
analyze the phonons in more complicated crystal structures.

Having simplified the interaction by making it pointlike,
and simplifed the ansatz by assuming the condensate varies
like a plane wave, in Ref. \cite{BowersLOFF} we
give the ansatz for the LOFF wave function,
and by variation obtain a gap equation which allows
us to solve for the gap parameter $\Delta$, the free energy and
the values of the diquark condensates which characterize
the LOFF state at a given $\delta\mu$ and $|{\bf q}|$. 
We then vary $|{\bf q}|$, to find the preferred (lowest
free energy) LOFF state at a given $\delta\mu$, and compare
the free energy of the LOFF state to that of the BCS state with
which it competes.\footnote{Our 
model Hamiltonian has two parameters, the four-fermion 
coupling $G$ and a cutoff $\Lambda$.  We 
often use the value of $\Delta_0$, the BCS gap 
obtained at $\delta\mu=0$, to 
describe the strength of the interaction: small $\Delta_0$
corresponds to small $G$.
When we wish to
study the dependence on the cutoff, we vary $\Lambda$
while at the same time varying the coupling $G$ such
that $\Delta_0$ is kept fixed.  We find that the relation
between other physical quantities and $\Delta_0$ is
reasonably insensitive to variation of $\Lambda$.}

Crystalline color superconductivity is favored for 
$\delta\mu_1<\delta\mu<\delta\mu_2$. As $\delta\mu$ increases,
one finds a first order phase transition from the ordinary
BCS phase to the crystalline color superconducting phase
at $\delta\mu=\delta\mu_1$ and then a second order
phase transition at $\delta\mu=\delta\mu_2$ at which $\Delta$          
decreases to zero.
Because the condensation
energy in the LOFF phase is much smaller than that of the BCS condensate
at $\delta\mu=0$, the value of $\delta\mu_1$ is almost identical
to that at which the naive unpairing transition from the 
BCS state to the state with no pairing would occur if
one ignored the possibility of a LOFF phase, 
namely $\delta\mu_1=\Delta_0/\sqrt{2}$.  For all practical
purposes, therefore, the LOFF gap equation is not required in order
to determine $\delta\mu_1$. The LOFF gap equation
is used to determine $\delta\mu_2$
and the properties of the crystalline color
superconducting phase \cite{BowersLOFF}.

We find that 
the LOFF gap parameter decreases from $0.23 \Delta_0$
at $\delta\mu=\delta\mu_1$
to zero at $\delta\mu=\delta\mu_2$~\cite{BowersLOFF}.  
The critical
temperature above which the LOFF state melts is 
$T_c = \Delta\sqrt{3/2\pi^2}$~\cite{Takada2,ngloff},
which means that, except for very close to $\delta\mu_2$,
$T_c$  will be much
higher than typical neutron star temperatures. 

In the limit of a weak four-fermion interaction, $G\rightarrow 0$,
the crystalline color superconductivity window is bounded by
$\delta\mu_1=\Delta_0/\sqrt{2}$ and $\delta\mu_2=0.754\Delta_0$,
as first demonstrated in Refs. \cite{LO,FF}.  These results have
been extended beyond the $G\rightarrow 0$ limit in Ref. \cite{BowersLOFF}.
Note that the BCS gap $\Delta_0$ increases monotonically with $G$.
We may therefore use $\Delta_0$ to parametrize the strength of 
the interaction $G$. This proves convenient because, as we have seen,
both $\delta\mu_1$ and $\delta\mu_2$ are given simply
in terms of the physical quantity 
$\Delta_0$. (Writing them in terms of the model-dependent
parameters $G$ and $\Lambda$ requires unwieldy expressions.)
As first recognized by Refs. \cite{LO,FF},
at any fixed $\delta\mu$ the LOFF phase only occurs when
the interaction strength (that is, $\Delta_0$) lies
in a specified window.  LOFF pairing does not survive the 
weak-coupling
($\Delta_0\rightarrow 0$) limit at fixed $\delta\mu$, because 
in this limit the width of the band on the Fermi surface 
in which pairing occurs goes to zero.  On the other hand,
if one takes  $\delta\mu$ and $\Delta_0$ both to zero
while keeping $\delta\mu/\Delta_0$ fixed and in the appropriate
range, LOFF pairing persists down to arbitrarily weak 
coupling~\cite{LO,FF,BowersLOFF}.

\section{Opening the Crystalline Color Superconductivity Window}

The first part of this section is more 
technical than the rest of this paper. Some readers
may wish to skip to the text after Eqs. (\ref{GapEq},\ref{Cdefs}).

The variational derivation of the gap equation for the crystalline
color superconducting phase is somewhat cumbersome \cite{BowersLOFF}.
One constructs a variational ansatz in which only quarks within
a ``pairing region'' are allowed to pair, minimizes the
free energy with respect to all variational parameters (two per mode
in momentum, color, flavor and spin space), and obtains
a self-consistency relation which may then be solved to obtain
$\Delta$.  The intricacy arises from the fact that the definition
of the boundary of the pairing region involves $\Delta$ itself.
In Ref. \cite{ngloff}, we provide a diagrammatic
rederivation in which one simply makes an ansatz for the 
quantum numbers of the condensate and then ``turns a field-theoretical
crank'' and sees this intricate result emerge.
We then use the diagrammatic derivation
to analyze the LOFF phase at nonzero temperature,
obtaining the result for $T_c$ given in the previous section.
The diagrammatic formalism also allows us to go beyond
a point-like interaction, and
treat the exchange of a propagating gluon~\cite{pertloff}.
Before presenting the results of this generalization
of the interaction, we first
sketch the diagrammatic derivation of the gap equation.

In the crystalline color superconducting phase,
the condensate contains pairs
of $u$ and $d$ quarks with momenta such that the total momentum
of each Cooper pair is given by $2{\bf q}$, with the direction
of ${\bf q}$ chosen spontaneously.  
As noted above, wherever there is an instability
towards
(\ref{simplifiedcondensate}), we expect the true
ground state to be a crystalline condensate which varies in space
like a sum of several such plane waves with the same $|{\bf q}|$.

In order to describe pairing between $u$ quarks with momentum 
${\bf p} + {\bf q}$ and $d$ quarks with momentum ${\bf -p} + {\bf q}$,
we must use a modified Nambu-Gorkov spinor defined as
\begin{equation} \label{Psi}
\Psi(p,q) = \left(\begin{array}{l} \psi_u(p+q) \\ \psi_d(p-q) \\
\bar\psi^T_d(-p+q) \\ \bar\psi^T_u(-p-q) \end{array}\right)\ .
\end{equation}
Note that by $q$ we mean the four-vector $(0,{\bf q})$. The
Cooper pairs have nonzero total momentum, and the ground state
condensate (\ref{simplifiedcondensate}) is static.
The momentum dependence of (\ref{Psi})
is motivated by the fact that in the presence of
a crystalline color superconducting condensate,
anomalous propagation does not only mean picking up or
losing two quarks from the condensate. It also means picking
up or losing momentum $2{\bf q}$.
The basis (\ref{Psi}) has been chosen so that the 
inverse fermion propagator in the crystalline
color superconducting phase is diagonal in $p$-space
and is given by
\begin{equation} \label{Sinv}
S^{-1}(p,q) = \left[\begin{array}{cccc} \slash{p}+\slash{q}+\mu_u \gamma_0
& 0 & -\bar{\Delta}(p,-q) & 0 \\ 0 & \slash{p}-\slash{q}+\mu_d
\gamma_0 & 0 & \bar{\Delta}(p,q) \\ -{\Delta}(p,-q) & 0 &
(\slash{p}-\slash{q}-\mu_d \gamma_0)^T & 0 \\ 0 & {\Delta}(p,q) 
& 0 & (\slash{p}+\slash{q}-\mu_u \gamma_0)^T \end{array}\right]\ ,
\end{equation}
where $\bar{\Delta} = \gamma_0 {\Delta}^{\dagger} \gamma_0$
and, here, ${\Delta}$ is a matrix 
proportional to $C\gamma_5\epsilon^{\alpha\beta 3}$. 
Note that the condensate
is explicitly antisymmetric in flavor.
$2{\bf p}$ is the relative momentum
of the quarks in a given pair and is different for different
pairs. In the gap equation below, we shall integrate over $p_0$
and ${\bf p}$. As desired, the off-diagonal blocks describe
anomalous propagation in the presence of a condensate of
diquarks with momentum $2{\bf q}$. 

We obtain the gap equation by solving
the one-loop Schwinger-Dyson equation given by
\begin{equation} \label{SDeq}
 S^{-1}(k,q)-S_0^{-1}(k,q) = -g^2 \int \frac{d^4p}{(2\pi)^4}
 \Gamma_\mu^A S(p,q)\Gamma_\nu^B D_{AB}^{\mu\nu}(k-p),
\end{equation}
where  $D_{AB}^{\mu\nu} = D^{\mu\nu} \delta_{AB}$ is the gluon
propagator, $S$ is the full quark propagator, whose inverse is 
given by (\ref{Sinv}), and $S_0$ is the fermion
propagator in the absence of interaction, given by $S$ with
${\Delta}=0$. The
vertices are defined as follows: 
\begin{equation} \label{vertex}
\Gamma_\mu^A = \left(\begin{array}{cccc} \gamma_\mu\lambda^A/2 & 0 & 0
& 0 \\ 0 & \gamma_\mu\lambda^A/2 & 0 & 0 \\ 0 & 0 &
-(\gamma_\mu\lambda^A/2)^T & 0 \\ 0 & 0 & 0 &
-(\gamma_\mu\lambda^A/2)^T \end{array}\right) .
\end{equation}

In Refs. \cite{BowersLOFF,ngloff}, we introduce a point-like
interaction by replacing $g^2 D^{\mu\nu}$ by
$g^{\mu\nu}$ times a constant $G$. 
After some algebra (essentially the determination of $S$ given
$S^{-1}$ specified above), and upon suitable projection,
the Schwinger-Dyson equation (\ref{SDeq}) reduces
to a gap equation for the gap parameter $\Delta$ given
(in Euclidean space) by
\begin{equation} \label{gapeqdelall}
\Delta = 2 G \int \frac{d^4p}{(2\pi)^4} \frac{4 \Delta w}{w^2 - 4
\left[ (|{\bf p}|^2 - (i p_0+\delta\mu)^2) ({\mu}^2 - |{\bf q}|^2) + 
({\bf p}\cdot{\bf q} + \mu (i p_0 + \delta\mu))^2 \right]}
\end{equation}
where $w=|{\bf p}|^2 - |{\bf q}|^2 - (i p_0+\delta\mu)^2 + {\mu}^2
+ \Delta^2$.
We show in Ref. \cite{ngloff} that upon neglecting the (small)
contributions of antiparticle pairing, this gap equation
simplifies to 
\begin{equation} \label{gapeqdel1}
\Delta = 2 G \int \frac{d^4p}{(2\pi)^4} \frac{2 \Delta
\sin^2\frac{\beta}{2}}{\left(p_0-i E_1({\bf p})\right)
\left(p_0+i E_2({\bf p})\right)} 
\end{equation}
where $E_{1,2}({\bf p})$ are defined as in Ref. \cite{BowersLOFF}:
\begin{eqnarray} \label{E12}
E_1({\bf p}) = & + \delta\mu + \frac{1}{2}
\left(|{\bf p}+{\bf q}|-|{\bf p}-{\bf q}|\right) + \frac{1}{2}
\sqrt{\left(|{\bf p}+{\bf q}|+|{\bf p}-{\bf q}|-2{\mu}\right)^2 +
4 \Delta^2 \sin^2\frac{\beta}{2}} \nonumber \\
E_2({\bf p}) = & -\delta\mu - \frac{1}{2}
\left(|{\bf p}+{\bf q}|-|{\bf p}-{\bf q}|\right) + \frac{1}{2}
\sqrt{\left(|{\bf p}+{\bf q}|+|{\bf p}-{\bf q}|-2{\mu}\right)^2 +
4 \Delta^2 \sin^2\frac{\beta}{2}}
\end{eqnarray}
and $\beta$ is defined as the angle between the up quark momentum
${\bf q}+{\bf p}$ and the down quark momentum ${\bf q}-{\bf p}$.
Upon doing the $p_0$ integration, we obtain the 
gap equation derived variationally in Ref. \cite{BowersLOFF}:
\begin{eqnarray} \label{ABRgapeq}
1 &=& 2 G \int_{{\bf p} \in {\cal P}}  \frac{d^3p}{(2\pi)^3} 
\frac{2\sin^2\frac{\beta}{2}}{E_1({\bf p})
+E_2({\bf p})}\nonumber\\
&=& 2 G \int_{{\bf p} \in {\cal P}}  \frac{d^3p}{(2\pi)^3} \frac{2
\sin^2\frac{\beta}{2}}{\sqrt{\left(|{\bf p}+{\bf q}| +
|{\bf p}-{\bf q}|-2{\mu}\right)^2 + 4 \Delta^2
\sin^2\frac{\beta}{2}}}
\end{eqnarray}
where the ``pairing region'' ${\cal P}$ in ${\bf p}$-space is
given by
\begin{equation} \label{Preg}
{\cal P} = \{ {\bf p} \ | \ E_1({\bf p}) > 0 \ {\rm and} \ E_2({\bf p}) >
0 \} \ .
\end{equation}
Thus, an exercise in residue calculus has reproduced
the blocking regions, excluding from the gap equation
those 
regions in momentum
space where $E_1({\bf p})$ or $E_2({\bf p})$ is negative.
Note 
that because $E_1({\bf p}) + E_2({\bf p})\geq 0$,
as can be seen from the definitions (\ref{E12}), there is no value of
${\bf p}$ for which both $E_1$ and $E_2$ are negative.
Note also that the 
gap equation is dominated by those regions in momentum
space where $E_1({\bf p})+E_2({\bf p})$ is as small
as possible, where the integrand in (\ref{ABRgapeq}) 
is of order $1/\Delta$.
These values of ${\bf p}$ are such that both members
of a LOFF pair have momenta close to (within $\sim \Delta$ of) 
their respective Fermi surfaces. That is, $|{\bf p+q}|$ is within
$\Delta$ of $\mu_u$ and $|{\bf -p+q}|$ is within $\Delta$ of $\mu_d$.
The results described in the previous section all follow
from analysis of the gap 
equation (\ref{ABRgapeq})~\cite{LO,FF,Takada2,BowersLOFF}.

In Ref. \cite{pertloff}, we begin with the Schwinger-Dyson
equation (\ref{SDeq}) but this time keep the gluon propagator.
That is, we analyze the crystalline color superconducting
phase upon assuming that quarks interact by the
exchange of a medium-modified gluon, as is
quantitatively valid at asymptotically high densities.
The medium-modified gluon propagator is given by
\begin{equation} \label{gluonprop}
D_{\mu\nu}(p) = {P^T_{\mu\nu}\over p^2-G(p)} + {P^L_{\mu\nu}\over p^2-F(p)} -
\xi {p_\mu p_\nu \over p^4}\ ,
\end{equation}
where $\xi$ is the gauge parameter, 
$G(p)$ and $F(p)$ are functions of $p_0$ and $|\bf p|$, and the
projectors $P^{T,L}_{\mu\nu}$ are defined as follows:
\begin{equation} \label{PLT}
P^T_{ij} = \delta_{ij} - \hat{p_i}\hat{p_j}, \ P^T_{00}=P^T_{0i}=0, \
P^L_{\mu\nu} = -g_{\mu\nu} + {p_\mu p_\nu \over p^2} - P^T_{\mu\nu}.
\end{equation}
The functions $F$ and $G$ describe the effects of the medium
on the gluon propagator.  If we neglect the Meissner effect (that is,
if we neglect the modification of $F(p)$ and $G(p)$ due to the gap
$\Delta$ in the fermion propagator) then $F(p)$ describes
Thomas-Fermi screening and $G(p)$ describes Landau damping and they are
given in the hard dense loop (HDL) approximation by \cite{LeBellac}
\begin{eqnarray} \label{GF}
F(p) &=& m^2 {p^2\over|{\bf p}|^2} \left( 1 - {ip_0\over|{\bf p}|} Q_0
\left( {ip_0\over|{\bf p}|} \right) \right), 
\ \ \ \ \ \  {\rm with}\ Q_0(x) = {1\over2} \log
\left( {x+1\over x-1} \right), \nonumber\\
G(p) &=& {1\over2} m^2 {ip_0\over|{\bf p}|} \left[ \left( 1 - \left(
{ip_0\over|{\bf p}|} \right)^2 \right) Q_0 \left( {ip_0\over|{\bf p}|}
\right) + {ip_0\over|{\bf p}|} \right] \ ,
\end{eqnarray}
where $m^2 = g^2 \mu^2/\pi^2$ is the Debye mass for $N_f=2$.
The further modification to the gluon propagator due
to the Meissner effect in spatially uniform color
superconducting phases has been the subject
of much recent work \cite{Meissner}, but
the Meissner effect in the crystalline color superconducting
phase has not yet been analyzed.  Fortunately,
in our calculation of $\delta\mu_2$ we
we shall only need to study the crystalline color
superconducting phase in
the limit in which $\Delta\rightarrow 0$, and in
this limit the expressions (\ref{gluonprop}) and
(\ref{GF}) are valid.

Upon neglecting the (small) contributions from antiparticle pairing,
the gap equation becomes
\begin{eqnarray} \label{GapEq}
\Delta(k_0) & = & \frac{- i g^2}{3 \sin^2{\frac{\beta(k,k)}{2}}} 
  \int \frac{d^4p}{(2\pi)^4}
\frac{\Delta(p_0)}{(p_0+E_1)(p_0-E_2)} 
\nonumber \\
 &&\times \left[ \frac{C_F}{(k-p)^2 - F(k-p)} + \frac{C_G}{(k-p)^2 - G(k-p)} + 
    \frac{C_\xi \xi}{(k-p)^2}\right],
\end{eqnarray}
where
\begin{eqnarray} \label{Cdefs}
C_F & = & \cos^2{\frac{\beta(k,p)}{2}}
 \cos^2{\frac{\beta(p,k)}{2}} 
 - \cos^2{\frac{\beta(k,-p)}{2}} \cos^2{\frac{\beta(-p,k)}{2}} -
 \sin^2{\frac{\beta(k,k)}{2}} 
 \sin^2{\frac{\beta(p,p)}{2}} ,
\nonumber \\
C_G  &=& \frac{\cos{\beta(k,-p)} \cos{\beta(-p,k)}-\cos{\beta(k,p)}
\cos{\beta(p,k)} }{2} -  2
\sin^2{\frac{\beta(k,k)}{2}} \sin^2{\frac{\beta(p,p)}{2}}
\nonumber \\
 && -\, \cos{\alpha(k,p)}\left(\cos{\alpha(p,k)} \sin^2{\frac{\beta(k,-p)}{2}}
+ \cos{\alpha(-p,-k)} \sin^2{\frac{\beta(k,p)}{2}}\right)
\nonumber \\
 && -\, \cos{\alpha(-k,-p)} \left(\cos{\alpha(p,k)}
\sin^2{\frac{\beta(p,k)}{2}} + \cos{\alpha(-p,-k)}
\sin^2{\frac{\beta(-p,k)}{2}}\right) , 
\nonumber \\
C_\xi & = & \sin^2{\frac{\beta(k,p)}{2}}
\sin^2{\frac{\beta(p,k)}{2}} - \sin^2{\frac{\beta(k,k)}{2}}
\sin^2{\frac{\beta(p,p)}{2}} - \sin^2{\frac{\beta(k,-p)}{2}}
\sin^2{\frac{\beta(-p,k)}{2}}\nonumber \\
 && +\, \cos{\alpha(k,p)} \left(\cos{\alpha(p,k)}
\sin^2{\frac{\beta(k,-p)}{2}} + \cos{\alpha(-p,-k)}
\sin^2{\frac{\beta(k,p)}{2}}\right) \nonumber \\
 && +\, \cos{\alpha(-k,-p)} \left(\cos{\alpha(p,k)}
\sin^2{\frac{\beta(p,k)}{2}} + \cos{\alpha(-p,-k)}
\sin^2{\frac{\beta(-p,k)}{2}}\right) ,
\end{eqnarray}
with $E_1$ and $E_2$ as in (\ref{E12}),
and where $\cos{\alpha(k,p)} = \widehat{(k-q)}\cdot\widehat{(k-p)}$
and $\cos{\beta(k,p)} = \widehat{(q+k)}\cdot\widehat{(q-p)}$.
In Ref. \cite{pertloff}, we use this gap equation to obtain
$\delta\mu_2$, the upper boundary of the crystalline
color superconductivity window. This analysis is controlled
at asymptotically high densities where the coupling $g$
is weak.  

At weak coupling,
quark-quark scattering by single-gluon exchange is
dominated by forward scattering.  In most scatterings, 
the angular positions of the quarks on their 
respective Fermi surfaces do
not change much.  As a consequence, the weaker the coupling
the more the physics can be thought of as a sum of many
$(1+1)$-dimensional theories, with only rare large-angle scatterings
able to connect one direction in momentum space with 
others~\cite{Hong}.  In the LOFF state, small-angle scattering
is advantageous because it cannot scatter a pair of
quarks out of the region of momentum space in which
both members of the pair are in their respective rings,
where pairing is favored.  
This means that
it is natural to expect that a forward-scattering-dominated
interaction like single-gluon
exchange is much more favorable for
crystalline color superconductivity than a point-like
interaction, which yields $s$-wave scattering.

Suppose for
a moment that we were analyzing a truly $(1+1)$-dimensional
theory.
The momentum-space geometry of the LOFF state in one spatial dimension
is qualitatively different from that in three.  
Instead of Fermi surfaces, we would have
only  ``Fermi points'' at $\pm \mu_u$ and $\pm \mu_d$.
The only choice of $|{\bf q}|$ which allows pairing between
$u$ and $d$ quarks at their respective Fermi points
is $|{\bf q}|=\delta\mu$.
In $(3+1)$ dimensions, in contrast, $|{\bf q}|>\delta\mu$ is
favored because it allows LOFF pairing in ring-shaped
regions of the Fermi surface, rather than 
just at antipodal points \cite{LO,FF,BowersLOFF}.
Also, in striking contrast to the $(3+1)$-dimensional case,
it has long been known that in a true $(1+1)$-dimensional 
theory with a point-like interaction between fermions,
$\delta\mu_2/\Delta_0\rightarrow\infty$ in the weak-interaction
limit \cite{LOFF1D}. 

We expect that in $(3+1)$-dimensional
QCD with the interaction given by single-gluon exchange,
as $\mu\rightarrow \infty$ and $g(\mu)\rightarrow 0$ 
the $(1+1)$-dimensional results should be approached: 
the energetically favored value of $|{\bf q}|$ should
become closer and closer to $\delta\mu$, and $\delta\mu_2/\Delta_0$
should diverge.  We derive both these effects in
Ref. \cite{pertloff}  and
furthermore show that both are clearly in evidence
already at the rather large coupling $g=3.43$, corresponding
to $\mu=400$ MeV using the conventions of Refs. \cite{SW3,Shuster}.
At this coupling, $\delta\mu_2/\Delta_0\approx 1.2$, 
meaning that $(\delta\mu_2-\delta\mu_1)\approx (1.2-1/\sqrt{2})\Delta_0$,
which is much
larger than $(0.754-1/\sqrt{2})\Delta_0$.
If we go to much higher densities, where the calculation
is under quantitative control, we find an even more striking
enhancement: when
$g=0.79$ we find $\delta\mu_2/\Delta_0 > 1000$!
We see that (relative to expectations based on experience
with point-like interactions)
the crystalline color superconductivity window
is wider by more than four orders of magnitude at this
weak coupling, and is about one order
of magnitude wider at accessible densities if weak-coupling results
are applied there.\footnote{LOFF
condensates have also recently been considered in two other contexts. 
In QCD with $\mu_u<0$, $\mu_d>0$ and $\mu_u=-\mu_d$, one has
equal Fermi momenta for $\bar u$ antiquarks and $d$ quarks,
BCS pairing occurs, and consequently a 
$\langle \bar u d\rangle$ condensate forms \cite{SonStephIsospin,Splittorff}.  
If $-\mu_u$ and $\mu_d$ differ,
and if the difference lies in the appropriate range, a LOFF
phase with a spatially varying $\langle \bar u d\rangle$
condensate results \cite{SonStephIsospin,Splittorff}.
The result of Ref. \cite{pertloff} that the LOFF window is much
wider than previously thought applies in this context also. 
Suitably isospin asymmetric nuclear matter may also admit LOFF pairing, as 
discussed recently in Ref. \cite{Sedrakian}.  Here, the interaction
is not forward-scattering
dominated.}

We have found that $\delta\mu_2/\Delta_0$ 
diverges in QCD as the weak-coupling, high-density limit
is taken.  
Applying results valid at asymptotically
high densities to those of interest in
compact stars, namely $\mu\sim 400$ MeV, we find
that even here the crystalline color superconductivity
window is an order of magnitude wider than that obtained
previously 
upon approximating the interaction between quarks as point-like.
The crystalline color superconductivity window in parameter
space may therefore be much wider than previously thought,
making this phase a {\it generic} feature of the phase diagram
for cold dense quark matter.  The reason
for this qualitative increase in $\delta\mu_2$ can
be traced back to the fact that gluon exchange 
at weaker and weaker coupling 
is more and more dominated by forward-scattering, while
point-like interactions describe 
$s$-wave scattering.  What is perhaps surprising is that even
at quite {\it large} 
values of $g$, gluon exchange yields 
an order of magnitude increase in $\delta\mu_2-\delta\mu_1$.

This discovery has significant implications for the QCD
phase diagram and may have significant
implications for compact stars.  At high enough baryon 
density
the CFL phase in which all quarks pair to form
a spatially 
uniform BCS condensate is favored. Suppose that as
the density is lowered the nonzero strange quark
mass induces the formation
of some less symmetrically paired quark matter
before the density is lowered so much that baryonic
matter is obtained. In this less symmetric quark matter,
some quarks may yet form a BCS condensate.  Those which
do not, however, will have differing Fermi momenta.
These will form a crystalline color superconducting
phase if the differences between their Fermi momenta
lie within the appropriate window.
In QCD, the interaction between quarks
is forward-scattering dominated and the 
crystalline
color superconductivity window is consequently
wide open. This phase is therefore generic,
occurring almost anywhere there 
are some quarks which cannot form BCS pairs.  
Evaluating
the critical temperature $T_c$ above which the crystalline
condensate melts requires solving the 
nonzero temperature gap equation obtained
from \eqref{GapEq} as done in Ref. \cite{ngloff} for
the case of a point-like interaction. As in that case, 
we expect that all compact stars which are minutes old or
older are much colder than $T_c$.  This suggests that wherever
quark matter which is not in the CFL phase occurs 
within a compact star, crystalline color superconductivity
is to be found. As we discuss in the next section,
wherever crystalline color superconductivity is found
rotational vortices may be pinned
resulting in the generation of glitches as the star
spins down.

Solidifying the implications of our results requires
further work in several directions.  First, we must
confirm that pushing Fermi surfaces apart via
quark mass differences has the same effect as pushing
them apart via a $\delta\mu$ introduced by hand.
Second, we must extend the analysis to the three
flavor theory of interest.  
And, third,
before evaluating the pinning force on a rotational
vortex and making predictions for glitch phenomena,
we need to understand which crystal structure is favored.

\section{Glitches in Compact Stars}

We do not yet know whether compact stars feature quark
matter cores.  And, we do not yet know whether, if they
contain quark matter,
that quark matter is color-flavor locked, meaning
that quarks of all colors and flavors participate
in BCS pairing, or whether the BCS condensate leaves
some quarks unpaired.  The lesson we take from the 
toy model analysis is that because the interaction 
between quarks in QCD
is dominated by forward scattering, rather than being
an $s$-wave point-like interaction, the difference in
Fermi momenta between the unpaired quarks need not
fall within a narrow window in order for them to form
a crystalline color superconductor.  

We wish now to ask whether the presence of a shell
of crystalline color superconducting quark matter 
in a compact star (between
the hadronic ``mantle'' and the CFL ``inner core'')
has observable consequences.  A quantitative formulation
of this question would allow one either to discover
crystalline color superconductivity, or to rule out
its presence.  (The latter would imply either no quark
matter at all, or a single CFL-nuclear interface~\cite{ARRW}.) 

Many pulsars have been observed to glitch.  Glitches are sudden
jumps in rotation frequency $\Omega$ which may
be as large as $\Delta\Omega/\Omega\sim 10^{-6}$, but may also
be several orders of magnitude smaller. The frequency of observed
glitches is statistically consistent with the hypothesis that 
all radio pulsars experience glitches~\cite{AlparHo}.
Glitches are thought to originate from interactions
between the rigid neutron star crust, typically somewhat 
more than a kilometer thick, and rotational vortices in a
neutron superfluid. 
The inner kilometer of crust
consists of a crystal lattice of nuclei immersed in 
a neutron superfluid~\cite{NegeleVautherin}.
Because the pulsar is spinning, the neutron superfluid 
(both within the inner crust and deeper inside the star) 
is threaded with
a regular array of rotational vortices.  As the pulsar's spin
gradually slows,
these vortices must gradually move outwards since the rotation frequency
of a superfluid is proportional to the density of vortices. 
Deep within the star, the vortices are free to move outwards.
In the crust, however, the vortices are pinned by their interaction
with the nuclear lattice.  
Models~\cite{GlitchModels} differ
in important respects as to how the stress associated
with pinned vortices is released in a glitch: for example,
the vortices may break and rearrange the crust, or a cluster
of vortices may suddenly overcome the pinning force and 
move macroscopically outward, with
the sudden decrease in the angular momentum
of the superfluid within the crust resulting in a sudden increase
in angular momentum of the rigid crust itself and hence a glitch.
All the models agree that the fundamental requirements
are the presence of rotational vortices in a superfluid 
and the presence
of a rigid structure which impedes the motion of vortices and
which encompasses enough of the volume of the pulsar to contribute
significantly to the total moment of inertia.

Although it is 
premature to draw quantitative conclusions,
it is interesting to speculate that some glitches may originate 
deep within a pulsar which features
a quark matter core, in a region of that core 
which is in
the crystalline color superconductor phase.
A full three-flavor analysis is required, first of all
in order to check whether the qualitative conclusions 
reached in the two-flavor analyses done to date
persist, and second of all in order
to determine whether the LOFF
phase is a superfluid.   If the only pairing is between $u$
and $d$ quarks, this 2SC phase is not a superfluid~\cite{ABR2+1},
whereas if the LOFF pairing involves the $s$ quarks, as
seems likely,
a superfluid is
obtained~\cite{CFL,ABR2+1}.
Henceforth, we suppose  that the LOFF phase is a superfluid, 
which means that if it occurs within a pulsar it will be threaded
by an array of rotational vortices.
It is reasonable to expect that these vortices will
be pinned in a LOFF crystal, in which the
diquark condensate varies periodically in space.
The diquark condensate vanishes at the core of a rotational
vortex, and for this reason the vortices will prefer
to be located with their cores pinned to the nodes of
the LOFF crystal.

A real calculation of the pinning force experienced by a vortex in a
crystalline color superconductor must await the determination of the
crystal structure of the LOFF phase. We can, however, attempt an order
of magnitude estimate along the same lines as that done by Anderson
and Itoh~\cite{AndersonItoh} for neutron vortices in the inner crust
of a neutron star. In that context, this estimate has since been made
quantitative~\cite{Alpar77,AAPS3,GlitchModels}.  
For one specific choice of parameters~\cite{BowersLOFF}, the LOFF phase
is favored over the normal phase by a free energy 
$F_{\rm LOFF}\sim 5 \times (10 {\rm ~MeV})^4$ 
and the spacing between nodes in the LOFF
crystal is $b=\pi/(2|{\bf q}|)\sim 9$ fm.
The thickness of a rotational vortex is
given by the correlation length $\xi\sim 1/\Delta \sim 25$ fm.  
The pinning energy
is the difference between the energy of a section of vortex of length 
$b$ which is centered on a node of the LOFF crystal vs. one which
is centered on a maximum of the LOFF crystal. It 
is of order $E_p \sim F_{\rm LOFF}\, b^3 \sim 4 {\rm \ MeV}$.
The resulting pinning force per unit length of vortex is of order
$f_p \sim E_p/b^2 \sim  (4 {\rm \ MeV})/(80 {\rm \ fm}^2)$.
A complete calculation will be challenging because
$b<\xi$, and is likely to yield an $f_p$
which is somewhat less than that we have obtained by dimensional 
analysis.
Note that our estimate of $f_p$ is
quite uncertain both because it is
only based on dimensional analysis and because the values
of $\Delta$, $b$ and $F_{\rm LOFF}$ are 
uncertain.  (We have a reasonable understanding of 
all the ratios $\Delta/\Delta_0$, $\delta\mu/\Delta_0$, $q/\Delta_0$ 
and consequently $b\Delta_0$ in the LOFF phase.  It is 
of course the value of the BCS gap $\Delta_0$ which is uncertain.) 
It is premature to compare our crude result 
to the results of serious calculations 
of the pinning of crustal neutron vortices as in 
Refs.~\cite{Alpar77,AAPS3,GlitchModels}.  It is nevertheless
remarkable that they prove to be similar: the pinning
energy of neutron vortices in the inner crust 
is $E_p \approx 1-3  {\rm \ MeV}$
and the pinning force per unit length is
$f_p\approx(1-3 {\rm ~MeV})/(200-400 {\rm ~fm}^2)$.


A quantitative theory of glitches originating within
quark matter in a LOFF phase must await further 
calculations, in particular a three flavor analysis and
the determination of the crystal structure of the QCD LOFF phase.
However, our rough estimate of the pinning force on 
rotational vortices in a LOFF region suggests that this force may be 
comparable to that on vortices in the inner crust of a conventional
neutron star.
Perhaps, therefore, glitches occurring in a region of crystalline
color superconducting quark matter may yield similar phenomenology
to that observed.  Were this to happen, we can hope
that a more detailed analysis
would reveal distinctions among observed glitches, with some
better understood as 
conventional glitches, originating in the inner crust,
and others better understood as
glitches originating deep within the star, in quark matter
in the crystalline color superconductor phase.
This is surely
strong motivation for further investigation.

\begin{theacknowledgments}
I am grateful to my collaborators, Mark Alford, Jeff Bowers, Joydip Kundu,
Adam Leibovich and Eugene Shuster, with whom I have been exploring
crystaline color superconductivity.  Let me also thank the
organizers of QCD@WORK for 
bringing together theorists and experimentalists who are putting QCD
to work in a variety of different arenas together for
a stimulating conference in a very
congenial setting.  This research was supported in 
part  by the DOE under cooperative research agreement \#DF-FC02-94ER40818
and through an OJI Award, and by the
A. P. Sloan Foundation.
\end{theacknowledgments}


\end{document}